\documentclass[a4paper]{article}
\usepackage{graphicx}
\usepackage{amssymb}
\usepackage{amsmath}

\begin{document}

\newcommand{\hdblarrow}{H\makebox[0.9ex][l]{$\downdownarrows$}-}
\title{A Method for Driving an Oscillator at a Quasi-Uniform Velocity}

\author{D.E. Zmeev$^{1,2}$\\
$^1$Department of Physics, Lancaster University,\\ Lancaster, UK\\
$^2$School of Physics and Astronomy, The University of Manchester,\\ Schuster Building, Oxford Road, Manchester, UK}

\date{12.07.2013}

\maketitle

\begin{abstract}

We describe a simple way to drive an actuator, comprising a superconducting coil moving in a static magnetic field, at a quasi-uniform velocity. The main objective is to avoid oscillations in this system with low damping, as they undermine the uniformity of the velocity. The method consists in calculating the force that should be exerted on the coil from the equation of motion and programming a waveform generator to produce the corresponding current through the coil. The method was tested on a device towing a grid through a closely fitted channel filled with superfluid $^4$He at temperatures below 100\,mK. The  motion of the grid over the distance of 4.3\,cm at 10\,cm/s resulted in oscillations of  less than 50\,$\mu$m in amplitude (or less than 1\,mm/s in terms of  velocity). The method can be applied to any oscillator.

PACS numbers: 47.27.-i, 67.25.dk, 84.50.+d, 03.65.Ge
\end{abstract}

\section{Introduction}
Homogeneous and isotropic turbulence is a much studied kind of turbulence both theoretically and experimentally in classical fluids. In superfluids in the low temperature limit, on the other hand, it is extremely difficult to produce. One way would be towing a grid at a uniform velocity through the superfluid contained in a channel. Such experiments have been done in superfluid $^4$He  at higher temperatures\cite{Stalp02}. The arising problem at low temperatures is that the actuators towing a grid must be frictionless, so as to not cause excess heating. Hence, oscillations occur during the motion of the actuator, which undermine the uniformity of the velocity. The unsuccessful attempts to drive such actuators (a linear superconducting motor\cite{Ihas06,Ihas08} and the ``floppy grid'' device\cite{Bradley11}) at a uniform velocity were described in earlier literature.

In this work we present a method for moving an electromagnetically driven actuator similar to the ``floppy grid'' device described in Ref.~4 at a quasi-uniform velocity. The motion consists of three stages: acceleration, uniform motion and deceleration. The acceleration and deceleration distances can be made very small, the limiting factor being the maximum electric current sustained by the actuator. The parameters responsible for the form of the time dependence of the driving force  are the effective resonant frequency and the effective damping parameter in the equation of motion of a simple harmonic oscillator with damping, which roughly governs the motion of the actuator.

\begin{figure}
\begin{center}
\includegraphics[%
  width=0.7\linewidth,
  keepaspectratio]{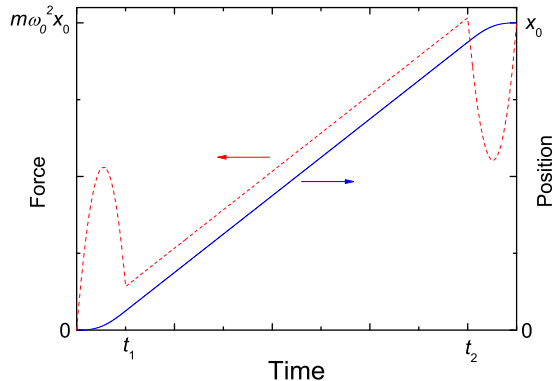}
\end{center}
\caption{(Color online) Examples of the calculated dependences of position $x(t)$ from (\ref{xt}) (solid blue  line) and the corresponding driving force $F(t)$ from (\ref{eq}) (dashed red  line).}
\label{force}
\end{figure}

\section{Theory}
Let us consider an oscillator whose motion follows the equation
\begin{equation}
F=m(\ddot x+\lambda \dot x + \omega_0^2 x),
\label{eq}
\end{equation}
where $F$ is the applied force, $m$ is the effective mass, $\lambda$ is the damping parameter
and $\omega_0$ is the resonant frequency in the absence of damping.

We are interested in driving such an actuator at a uniform velocity $v_0$ from one equilibrium position ($x=0$) to another ($x=x_0$) without inducing oscillations. If one applies a force linear in time, large oscillations occur inevitably\cite{Bradley11}.
We propose to shape the force in such a way that acceleration and deceleration take finite time, but the resulting motion is free from oscillations. For instance, the second derivative of the $x(t)$ dependence is continuous everywhere, so that the calculated force does not have discontinuities. One of the possibilities is the following dependence:
\begin{equation}
x(t) = \begin{cases} At^4+Bt^3, & 0 <t \leq t_1\\v_0 (t-t_0), & t_1<t\leq t_2\\ x_0-(A(t_3-t)^4+B(t_3-t)^3), & t_2<t\leq t_3; \end{cases}
\label{xt}
\end{equation}
where $t_3 = t_2 + t_1$. Parameters $A$, $B$ and $t_0$ are calculated using the conditions
of continuity of  $x$, $\dot x$ , and $\ddot x$  at $t = t_1$. The resulting motion is  not very different from the uniform motion.  The force required for such motion can be calculated by substituting
(2) in (1).  The typical shapes of the dependence of the force and position on time are shown in Fig.~\ref{force}. The acceleration and deceleration can
take arbitrarily small distances/times provided large enough forces can be sustained by the device. In practice, parameters $\lambda$ and $\omega_0$ can be measured independently or selected to result in motion with the smallest level of oscillations.

Acceleration profiles similar to (\ref{xt}) can be used to drive actuators governed by equations different from (\ref{eq}).

\begin{figure}
\begin{center}
\includegraphics[%
  width=0.7\linewidth,
  keepaspectratio]{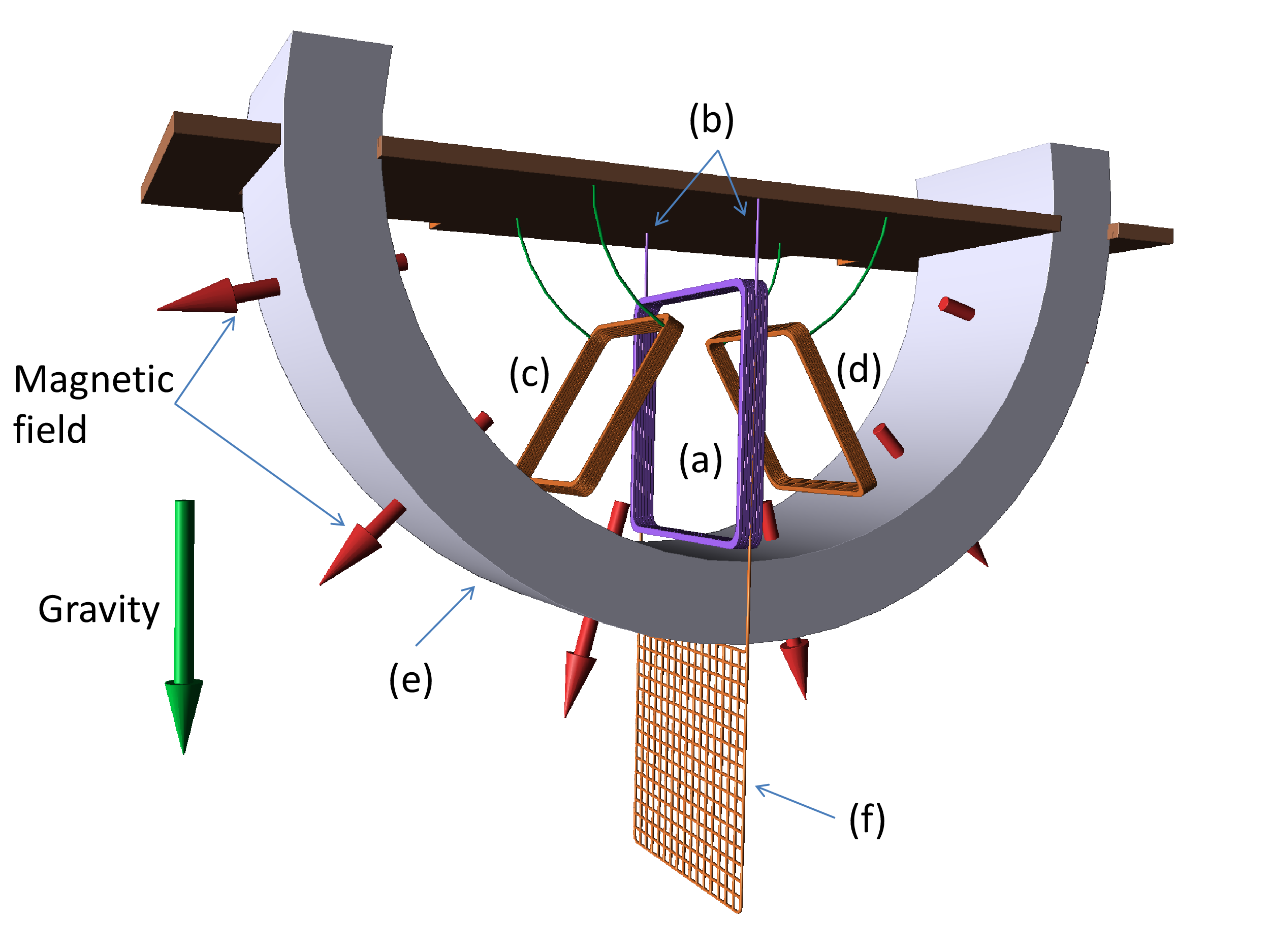}
\end{center}
\caption{(Color online) Experimental setup. The directions of the gravitational field and magnetic field in the experiment are indicated by green and red arrows respectively. The parts of the actuator are: driving coil (a) , elastic supports  (b), sensing coils  (c) and (d), ring magnet (e) and brass grid (f).}
\label{setup}
\end{figure}

\section{Experimental details}

The sketch of the device used in the experiment is presented in Fig.~\ref{setup}. It comprised a superconducting coil (a) (W $\times$ L $\times$ H = 15\,mm $\times$ 18\,mm $\times$ 2\,mm) wound from 40 turns (3 layers) of a single-filament NbTi wire clad in copper-nickel alloy (the wire diameter was 100\,$\mu$m). The two ends of the winding  $\approx$4\,mm in length served as  elastic supporting legs (b) and were fed through two holes in a circuit board and were secured by soldering. The bottom section of the coil was situated near the inner surface of a NdBFe ring magnet\cite{Magnet} (e) (O.D. $\times$ I.D.$ \times$ H = 5.5\,cm $\times$ 4.5\,cm $\times $1.5\,cm). The magnet was magnetized radially, i.e. the magnetic field was normal to the inner and outer surfaces of the magnet and the magnetic field on the surface was measured to be $\sim$0.1\,T at room temperature. When an electric current passes through the coil, it deflects from the vertical position and assumes a new equilibrium position defined by the magnitude of the current, stiffness of the supports and the gravitational force acting upon the coil and the square brass grid (f) (16\,mm$\times$16\,mm, 0.2\,mm thick, mesh size 0.7\,mm)  attached to the long sides of the coil. The current required for a $45^\circ$ deflection, which corresponds to a 2.5\,cm displacement of the middle point of the grid, was $\approx$0.6\,A~\---~far less than the critical current of the superconducting wire. In the actual experiment the grid moved inside a closely fitted curved channel with two 1.5\,mm wide circular slits to allow passing of the grid supports. The device with the channel resided inside a 0.6\,$\ell$ brass cell, which was  thermally anchored to the mixing chamber of the dilution refrigerator and could be filled with superfluid $^4$He. The resonant frequency of the device in vacuum at 30\,mK was found to be 4.8\,Hz and the quality factor was $\sim$2000.

The free ends of the main coil were spot-welded to a superconducting twisted pair passing via a hermetic feedthrough on the top of the experimental cell. The twisted pair had  good thermal contact with all stages of the dilution refrigerator and was soldered to a copper twisted pair at the 4.2\,K flange of the vacuum can. The driving current was generated by the Kepco BOP 100-1M bipolar power supply, which was controlled by the voltage produced by the arbitrary waveform generator Agilent 33522A. A small (100\,$\mu$A) 100\,kHz current was superimposed on the driving current, so that the position of the main coil could be monitored using  inductive coupling to two stationary superconducting sensing coils (c) and (d). The voltage across each coil was measured using two SRS SR830 lock-in amplifiers synchronized with the 100\,kHz source\cite{Bradley11,Polturak98}. The response of the coils was measured as the function of the position of the actuator at room temperature, and then at low temperatures the voltage across the two sensing coils was measured on a slow scan of the driving current. This allowed us to calibrate the position of the actuator as a function of the passed current. A noticeable hysterisis of $\sim$1\,mm linked to magnetic flux unpinning  was observed on slow scanning of the driving current in opposite directions.

\begin{figure}
\begin{center}
\includegraphics[%
  width=0.7\linewidth,
  keepaspectratio]{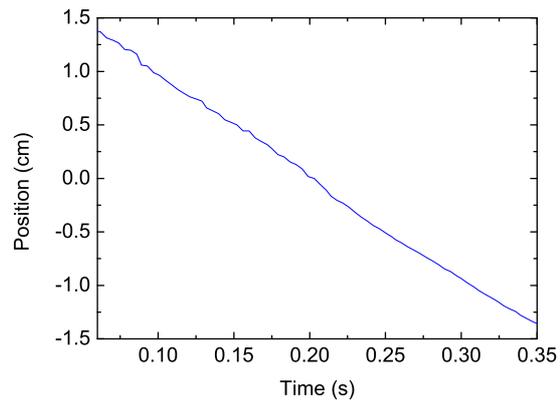}
\includegraphics[%
  width=0.7\linewidth,
  keepaspectratio]{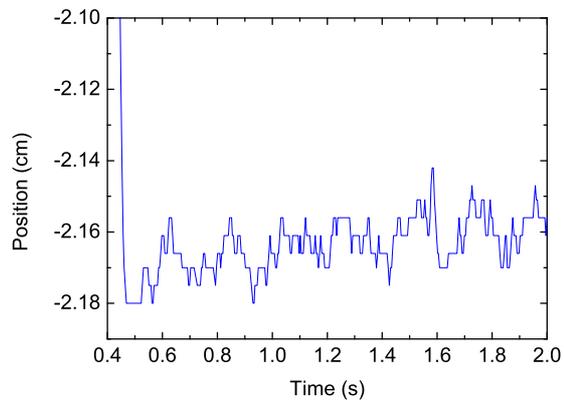}
\end{center}
\caption{(Color online) Examples of the recorded  position of the grid inferred from the signals from the two sensing coils. The driving pulse was applied at $t=0$\,s, and the grid moved from $x_0=$2.13\,cm to $x_1=$-2.17\,cm  in 0.43\,s. The temperature was 60\,mK and the pressure was 0.1\,bar.}

\label{position}
\end{figure}

\section{Results}
Since the actuator was moved over a wide angle (from $\approx +45^\circ$ to $\approx -45^\circ$, measured from the vertical position) and at high velocities, its actual equation of motion could have been very different from ~(\ref{eq}). Nevertheless, the effective values of the parameters $\omega_0$ and $\lambda$ in Eq.~\ref{eq} could have been selected to result in motion with a small level of oscillations. The criterion for minimisation was the amplitude of the oscillations immediately after stopping the actuator. The inferred position of the actuator as a function of time and an example of the oscillations after a 4.3\,cm stroke at a velocity of 10\,cm/s are presented in Fig.~\ref{position}. The oscillations during the motion were smaller than the uncertainty of the calibration of the actuator's position. The optimal values of the parameters did not depend on temperature, and therefore on properties of the superfluid $^4$He, up to 1.3\,K at a pressure of 0.1 bar. This fact can be explained by a large effective inertia of the device, which is hardly affected by the helium drag. This property proved indispensable and saved a lot of time in the experiments studying the temperature dependence of the decay of turbulence produced by the grid moving inside the channel\cite{Zmeev13}.

The motion of the actuator was highly reproducible and did not change noticeably after $\sim10^4$ strokes.

\section{Conclusion} We have developed a method for driving a generic oscillator at a quasi-uniform velocity without inducing oscillations. The method was applied to a device similar to the ``floppy grid''\cite{Bradley11}, and the motion of the grid through superfluid $^4$He at 10\,cm/s and for a stroke of 4.3\,cm resulted in oscillations of the grid at only $\sim$1\,mm/s.

\section{Acknowledgements}
The author would like to thank M.~Fear, P.~Walmsley and S.~Fisher for useful discussions of the design of the device.  This work was supported through the Materials World Network program by the Engineering and Physical Sciences Research Council [Grant No. EP/H04762X/1].

\end{document}